\documentclass[a4paper, 12 pt, fleqn]{article}

\usepackage{authblk}
\title{A theoretical framework and the experimental dataset for benchmarking numerical models of dilute pyroclastic density currents.}
\author[1]{Matteo Cerminara}
\author[2]{Ermanno Brosch}
\author[2]{Gert Lube}
\affil[1]{Istituto Nazionale di Geofisica e Vulcanologia, Pisa, Italy.}
\affil[2]{Volcanic Risk Solutions, Institute of Agriculture and Environment, Massey University, New Zealand.}
\date{November 13, 2018}
\usepackage{syntonly}

\usepackage{array}
\usepackage{bm}
\usepackage{bmpsize}
\usepackage{booktabs}
\usepackage[font = scriptsize, format=hang, labelfont = {sf, bf}]{caption}
\usepackage{chngpage}
\usepackage{enumitem}
\usepackage{eurosym}
\usepackage{fancyhdr}
\usepackage{float}
\usepackage{footmisc}
\usepackage{indentfirst}
\usepackage{layaureo}
\usepackage{listings}
\usepackage{longtable}
\usepackage{microtype}
\usepackage{mparhack}
\usepackage{showlabels}
\usepackage{siunitx}
\usepackage{subfig}
\usepackage{tabularx}
\usepackage{url}
\usepackage{xcolor}
\usepackage{wrapfig}
\usepackage{cancel} 
\usepackage{natbib} 

\usepackage[latin1]{inputenc}
\usepackage{lmodern}
\usepackage[T1]{fontenc}
\usepackage[english]{babel}

\usepackage{graphicx}

\usepackage{textcomp}
\usepackage{verbatim}
\usepackage{latexsym}
\usepackage{amsmath, amssymb, amsthm, eucal, multicol}
\usepackage{esdiff}
\usepackage{amsfonts}

\usepackage[pagebackref]{hyperref}
\usepackage[hyperpageref]{backref}
\renewcommand{\backref}[1]{}

\usepackage{multibib}
\newcites{web}{Web sites}

\begin{document}
\maketitle 
%

\section{Introduction.}
The aim of this document is to define a
Pyroclastic Density Currents (PDCs) benchmark based on a large-scale experiment to be used with numerical models at  different levels of complexity.
The document is organized as follows. 
Section~\ref{sect:exp} concisely describes the large-scale laboratory experiment setup and geometry, and the relevant specific bibliography.
Section~\ref{sect:mod} introduces the theoretical framework to adapt the experimental dataset to numerical models at different levels of complexity.
Section~\ref{sect:bc} details the initial and boundary conditions. In particular,
Section~\ref{sect:vel} describes the inlet velocity that best reproduces the experimental boundary conditions. 
Section~\ref{sect:temp} describes in detail the inlet concentration and temperature profiles, respectively.
Section~\ref{sect:gsd} describes the input grain size distribution.
Section~\ref{sect:postproc} gives the guidelines for consistently presenting the numerical outputs in a model inter-comparison study.
Section~\ref{sect:concl} is a summary to guide the reader through the data sets.

\section{Experimental Setup} \label{sect:exp}
The large-scale experiment that forms the base for this benchmark was conducted at the eruption simulator PELE \citep[the Pyroclastic flow Large-scale Experiment, fully described in ][]{lube2015synthesizing} in New Zealand, and simulates a fully dilute, fully turbulent PDC (i.e. a pyroclastic surge). 
PELE is a test facility to synthesize, view and measure inside dynamically scaled analogues of natural PDCs \citep{breard2016coupling,breard2017inside,breard2018enhanced}. Experimental currents of up to 6 tonnes of natural volcanic material and gas reach velocities of 7-32~m/s, flow thicknesses of 2-4.5 m, and runouts of >35 m. 

\begin{figure}[t!]
	\centering
	\includegraphics[width=\columnwidth]{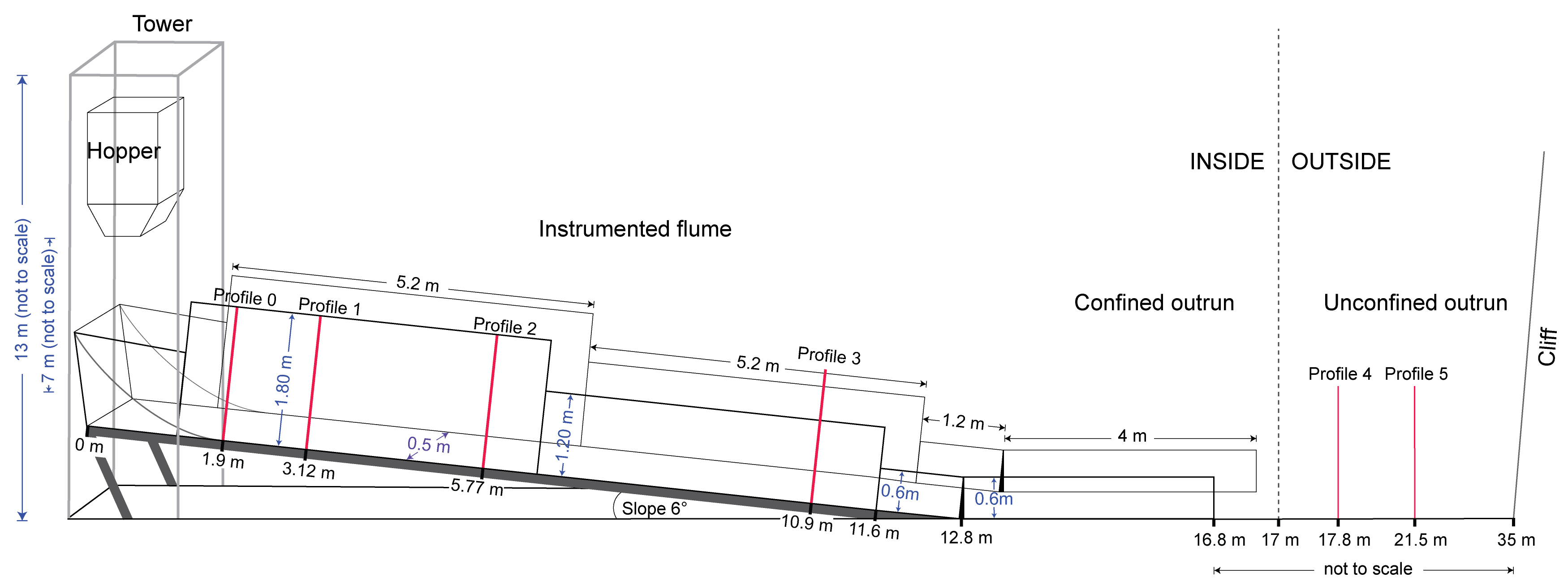}
	\caption{Sketch of the PELE set-up as configured for the benchmark experiment and outlining the main structural components of the facility: the Tower holding the heatable hopper, the inclined instrumented flume section, the horizontal, confined flume section and the unconfined outrun area.}
	\label{fig:PELEsetup}
\end{figure}

In the current experiment, an experimental pyroclastic density current is generated by the controlled gravitational collapse of a 120$^\circ$~C hot aerated suspension of natural volcanic particles (124~kg) and air from an elevated hopper into an instrumented flume. The mixture falls into and propagates through a 12.8~m long, 0.5~m wide and 6$^\circ$ inclined and instrumented flume, which is followed by a 4 m long, 0.5~m wide horizontal flume section leading onto a distal 18.2~m long unconfined outrun section (Fig.~\ref{fig:PELEsetup}). The sidewalls of the confined proximal and medial flume sections are composed of smooth high-temperature glass (on true right) and smooth steel (on true left). A mixture of sub-rounded, 4-8 mm diameter grey wacke pebbles was glued to the flow base to create a rough substrate with an average effective roughness of c. 5 mm. 

\begin{figure}[t!]
	\centering
	\includegraphics[width=\columnwidth]{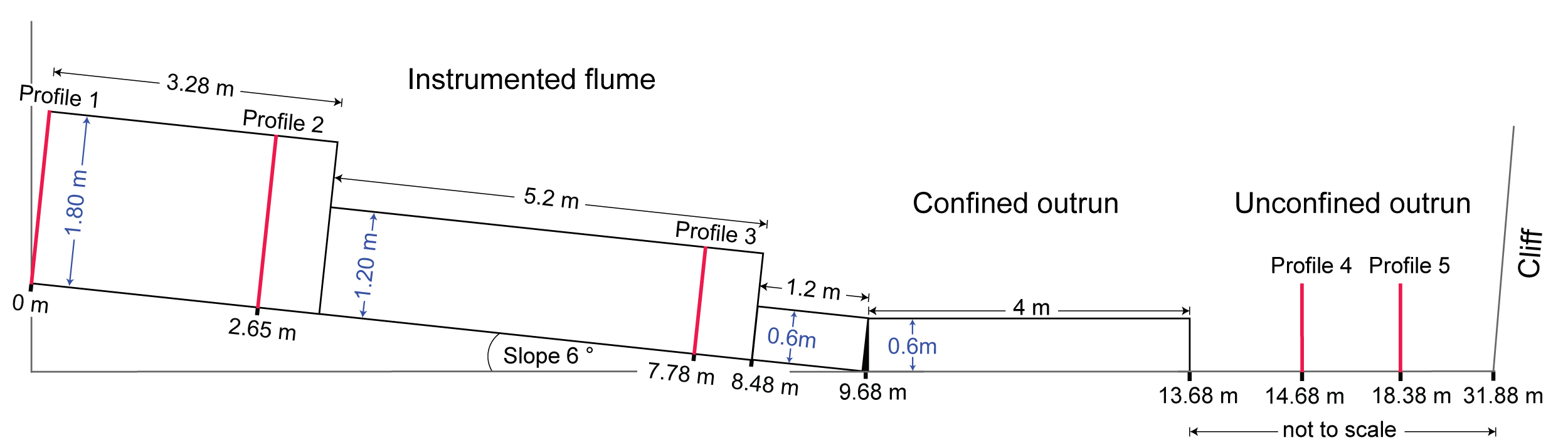}
	\caption{Sketch of the computational domain for the benchmark exercise starting at Profile 1. Note that the lengths noted at the base are distances measured along the flow runout starting at Profile 1.}
	\label{fig:PELEsetup2D}
\end{figure}

The flow is captured by 3 high-speed cameras, 11 normal speed cameras and 1 thermal infra-red camera viewing different flow regions during runout. Time-series of flow velocity, flow temperature, flow solid mass fraction, and flow grain-size distribution are obtained at orthogonal profiles at four different distances (Profiles 0, 1, 2 and 3 in Figures 1 and 2). Time-integrated mass and grain-size distributions are obtained at orthogonal profiles at six different distances (Profiles 0-5 in Figs.~\ref{fig:PELEsetup} and~\ref{fig:PELEsetup2D}). High-resolution pre- and post-experiment Terrestrial Laser Scan measurements are used to constrain space-variant geometry of the flow deposit. In addition, sediment samplers capture deposit mass per unit area at regular runout distances and are used to obtain downstream profiles of deposit grain-size distribution. 

In order to reduce the complexity of the benchmark exercise, the flow is only considered from Profile 1 at the inclined flume, i.e. complex processes of initial mixing with ambient air during free fall and impact with the channel are excluded from the computational domain. Fig.~\ref{fig:PELEsetup2D} depicts the geometry of computational domain from Profile 1 onwards.

The ambient conditions measured during the experiment here described are listed in Tab.~\ref{tab:ambient}.
\begin{table}[h!]
\centering
\begin{tabular}{ccc}
\toprule
$p$ [Pa] & $T_\mathrm{a}$ [$^\circ$ C] & humidity\\
\midrule
1.013e5 & 11.0 & 60\%\\
\bottomrule
\end{tabular}
\caption{Ambient conditions of the presented experiment.}
\label{tab:ambient}
\end{table}

\section{Model setup}\label{sect:mod}
In order to define a numerical case based on the PELE experiment, we need to prescribe boundary and initial conditions similar to the experimental settings. 

To allow the settings of boundary conditions for most of the models used for PDC simulations, we do not represent the initial source complexity and impose boundary conditions at the inlet of the inclined channel, perpendicular to the rough bottom wall. This is the most critical boundary for numerical simulations. We place this plane at Profile 1 of Fig. \ref{fig:PELEsetup}, 3.12 m ahead the beginning of the channel. 

As the flow can be considered as incompressible (its velocity is everywhere much lower than the speed of sound), the following fields at the inlet section completely define the injection of the multiphase gas-particle mixture into the channel section:
\begin{itemize}
\item parallel velocity field of the mixture: $u(z, t)$
\item orthogonal velocity field of the mixture: $v(z, t)$
\item parallel and orthogonal velocity fields of solid phases and gas component: $u_{j}(z,t)$, $v_{j}(z,t)$; $u_\mathrm{g}(z,t)$, $v_\mathrm{g}(z,t)$
\item temperature field of the mixture: $T(z,t)$
\item temperature fields of solid phases and gas component: $T_j(z,t)$; $T_\mathrm{g}(z,t)$
\item concentration or mass fraction fields of the solid phase: $\epsilon_\mathrm{s}(z,t)$, $y_\mathrm{s}(z,t)$
\item mass fraction fields of solid phases and gas component: $y_j(z,t)$; $y_\mathrm{g}(z,t)$
\end{itemize}

\subsection{Equilibrium assumption at the inlet}
Because PIV analysis and thermocouples do not yet allow us to measure the velocity and temperature of each individual particle class, we assume thermal and kinematic equilibrium between gas and particles at the inlet. 
Specifically, we assume:
\begin{itemize}
	\item kinematic equilibrium: $u_{j} = u_\mathrm{g}=u$, $v_{j}=v_\mathrm{g}=v$
	\item thermal equilibrium: $T_j = T_\mathrm{g} = T$
	\item we disregard particle settling velocity with respect to parallel velocity: $|u| \gg |v| \simeq 0$
\end{itemize}
This reduces the number of fields to be input at the inlet. 

\subsection{Inlet flow profiles}
The inlet flow profiles can be obtained from experimental data, and functions of $(z,t)$ are provided in this document in two forms:
\begin{itemize}
	\item By fitting data with appropriate functions of space and time (deriving from theoretical considerations);
	\item By using directly the measured raw data, after appropriate interpolation in space and time.
\end{itemize}
Our final choice is a mix of this two options: when we have a suitable theoretical fitting profile, we use it; otherwise we use interpolation techniques.

\subsection{Models/data hierarchy.}
Finally, the inlet flow functions can be used as boundary conditions for models at different levels of complexity:
\begin{itemize}
	\item Transient 2D/3D models: inlet fields are kept variable with height and time (variability in the y direction - orthogonal to the view plane - at the inlet is neglected in 3D models).
	\item Stationary 2D/3D models: inlet profiles are time-averaged, to keep only the spatial dependence, $f(z) = \langle f(z,t) \rangle_t$
	\item depth average transient 1D models: inlet profiles are spatially-averaged, to keep only the temporal dependence, $f(t) = \langle f(z,t) \rangle_z$
	\item depth average stationary 1D models: inlet profiles are averaged both in space and time
	\item other integral models: inlet mass, momentum and energy flow rates can be used
\end{itemize}

\section{Inlet boundary conditions}\label{sect:bc}
\subsection{Parallel velocity field.}\label{sect:vel}
The velocity profile can be split in two principal contributions $u(z,t) = \bar{u}(z,t) + u'(z,t)$, where:
\begin{enumerate}
	\item the mean profile, $\bar{u}(z,t)$, keeps the slowest time variability associated with unsteady mass release from the hopper;
	\item the velocity fluctuations, $u'(z,t)$, describe turbulent eddies at inlet.
\end{enumerate}
The procedure described hereafter allowed us to separate these two components and to obtain a coherent description of the subsequent flow field.

\subsubsection{The theoretical model for inlet velocity.}
The parallel velocity profile obtained from the PIV data has been fitted in space and time with a continuous and derivable function. 
This function is composed of two regions: 1) the inner layer, which is the region where the wall boundary layer develops; 2) the outer layer, the region above the boundary layer where a free-surface turbulent layer develops. Following the classic empirical results obtained in turbulence studies \citep[e.g., ][]{Altinakar1996}, we fit the inner layer with a power function, whereas the outer one with a Gaussian profile. 

At each time, the mean velocity profile is thus represented mathematically by:
\begin{equation}\label{eq:mean_velocity}
\bar{u}(z,t)=u_\mathrm{pg}(z, t) = u_\mathrm{pg}(\eta) = U \eta^{\xi} \exp\left[-\left(\frac{\eta -1 }{\chi}\right)^2 - \xi(\eta-1)\right]\,,
\end{equation}
where $\eta \equiv z/z_u$, so that height is made dimensionless with respect to $z_u$.
\begin{figure}[t!]
\centering
\includegraphics[width=\columnwidth]{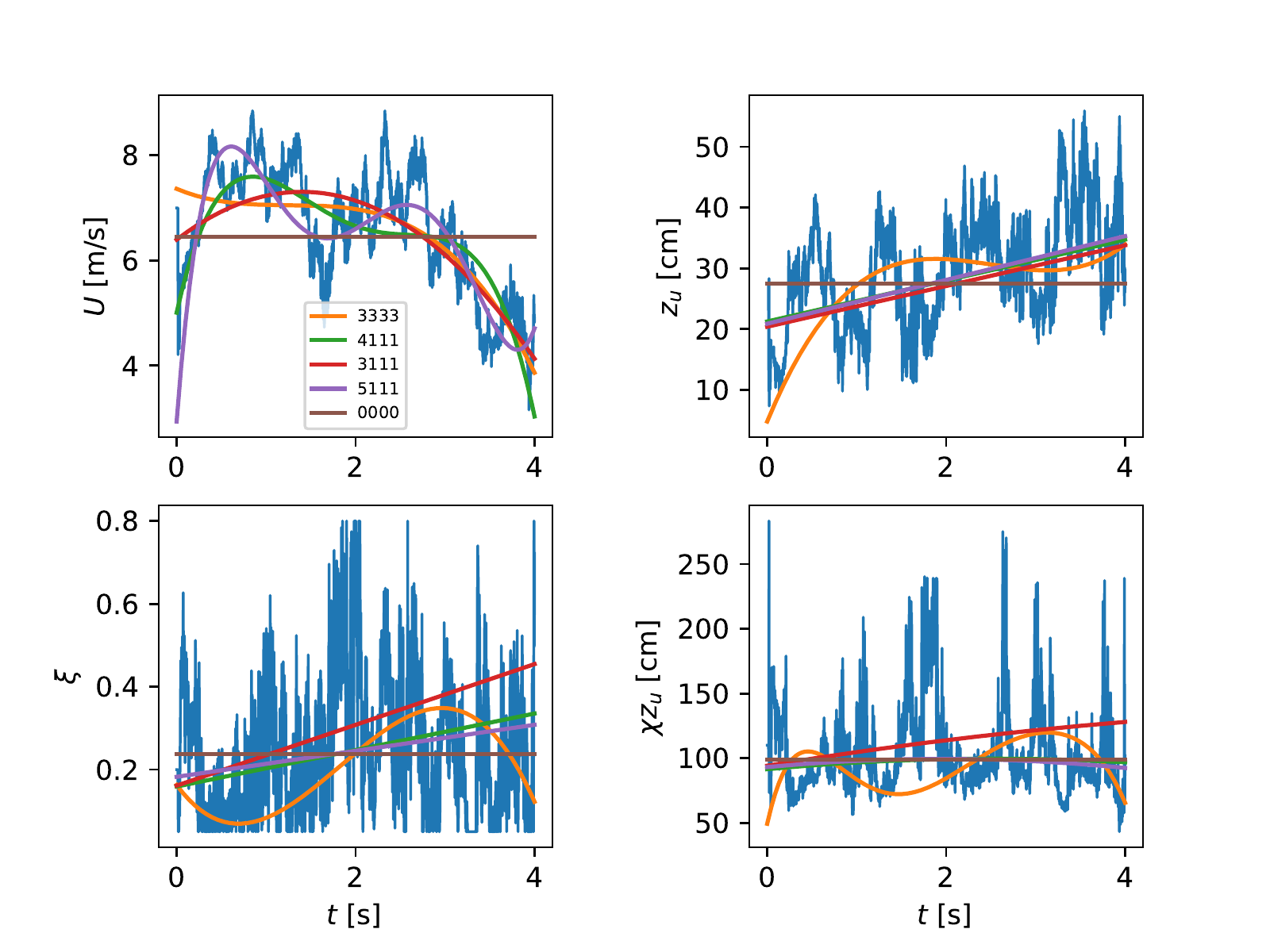}
\caption{Evolution of the four functions defining the mean velocity profile as defined in Tab.~\ref{tab:velocityMean}. The outer layer thickness is plotted in its dimensional form: $\chi z_u$. Blue lines are obtained from the fit without any temporal correlation, while orange, green, red, purple and brown are different polynomial fitting orders.}
\label{fig:par_pgt}
\end{figure}
This profile is a differentiable version of the profile proposed by \citet{Altinakar1996}, who proposed a power law in the boundary layer and a Gaussian profile in the outer layer, requesting continuity but not differentiability. 
A profile from Eq.~\eqref{eq:mean_velocity} is shown in Tab.~\ref{tab:velocityMean0000}. Here, $U,\, z_u\,, \xi\,, \chi$ all are function of time, representing the maximum velocity, its height, the inner layer exponent and the outer layer e-folding thickness. The latter can be considered as an effective thickness of the current.

\subsubsection{Fitting of time-dependent velocity data}
To obtain the temporal variability of the four parameters in Eq.~\ref{eq:mean_velocity}, we have applied a minimization algorithm for the residuals between $u_\mathrm{pg}(z, t)$ and the raw data $u(z,t)$.
In Fig.~\ref{fig:par_pgt} we show these four functions, as obtained by this procedure.  
\begin{figure}[t!]
\centering
\includegraphics[width=\columnwidth]{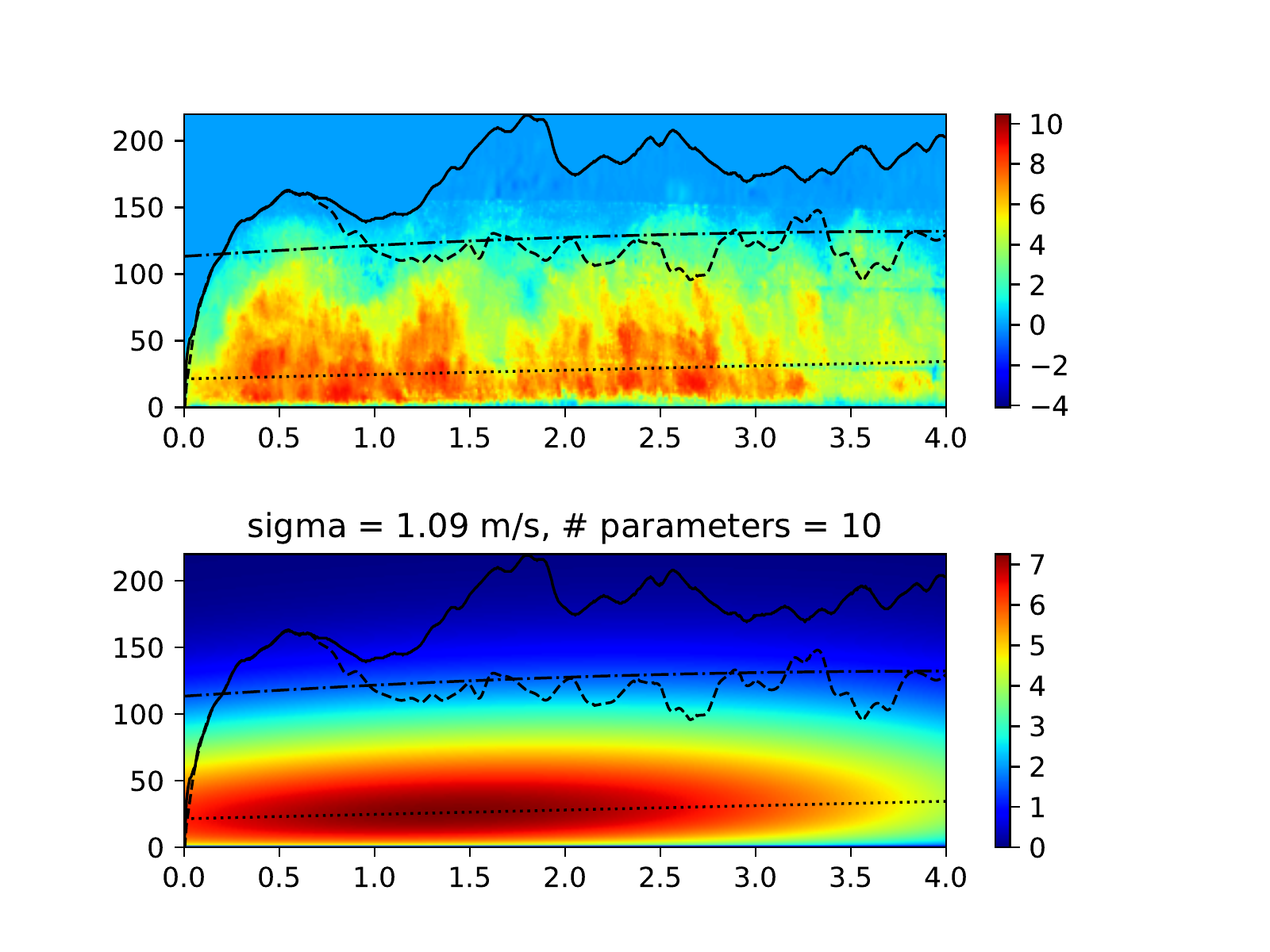}
\caption{Parallel flow velocity. Upper panel, raw data; lower panel, mean function $u_\mathrm{pg}$. Color scale is in m/s; abscissa is time in seconds, ordinate is height in centimeters. Relevant heights are plotted as a function of time: $h_\mathrm{max}$ solid line; $h_\mathrm{body}$ dashed line; $h$ dash-dot line; $z_u$ dotted line.}
\label{fig:par_mean}
\end{figure}
There, the blue lines represent the fitted values of $U,\, z_u\,, \xi\,, \chi$ without any temporal correlation. The other lines represent fitted values when we assume a polynomial form of the temporal evolution. Four cases are shown, where the polynomial order varies from 1st to 5th. 
The choice that minimizes the number of free parameters, keeping a satisfactory mean result, is a fit with polynomial degrees of (3,1,1,1) for $U,\, z_u\,, \xi\,, \chi$, respectively. The standard deviation in this case turns out to be $\sigma = 1.09$ m/s with 10 fitting parameters.  The best fitting parameters are reported in Tab.~\ref{tab:velocityMean}.
In Fig.~\ref{fig:par_mean}, we show the space-time contour plot of the raw PIV data and the fitting function $u_\mathrm{pg}(z,t)$. Moreover, in this figure we plot the inner layer thickness $z_u(t)$, the body height of the current as obtained from visual analysis of the PIV data $h_\mathrm{body}(t)$, the maximum height of the current again from visual analysis $h_\mathrm{max}$, and the current height as obtained from the e-folding thickness $h(t) = z_u(t) (\chi(t) + 1)$.

\begin{table}[h]
\centering
\begin{tabular}{cc}
\toprule
{\bf Transient, 2D/3D} & \\
\midrule
maximum velocity & $U(t) = U_0 + U_1 t + U_2 t^2 + U_3 t^3$\vspace{4pt}\\
$U_0$ & 6.378 m/s\\
$U_1$ & 1.267 m/s$^2$\\
$U_2$ & -0.4550 m/s$^3$\\
$U_3$ & -1.049e-4 m/s$^4$\\
\midrule
maximum velocity height & $z_u(t) = z_0 + z_1 t$\vspace{4pt}\\
$z_0$ & 21.42 cm\\
$z_1$ & 3.276 cm/s\\
\midrule
wall exponent & $\xi(t) = \xi_0 + \xi_1 t$\vspace{4pt}\\
$\xi_0$ & 0.1662\\
$\xi_1$ & 4.188e-2 s$^{-1}$\\
\midrule
outer layer thickness & $\chi(t) = \chi_0 + \chi_1 t$\vspace{4pt}\\
$\chi_0$ & 4.294\\
$\chi_1$ & -0.3661 s$^{-1}$\\
\bottomrule
\end{tabular}
\caption{The 10 parameters of the space- and time- fitting the raw velocity data. Temporal variability of the four parameters has been fitted with polynomials of degree 3,1,1,1.}
\label{tab:velocityMean}
\end{table}

\subsubsection{Time-averaged velocity data.}\label{sec:time_averaged_profiles}
\begin{figure}[h]
	\centering
	\includegraphics[width=0.7\columnwidth]{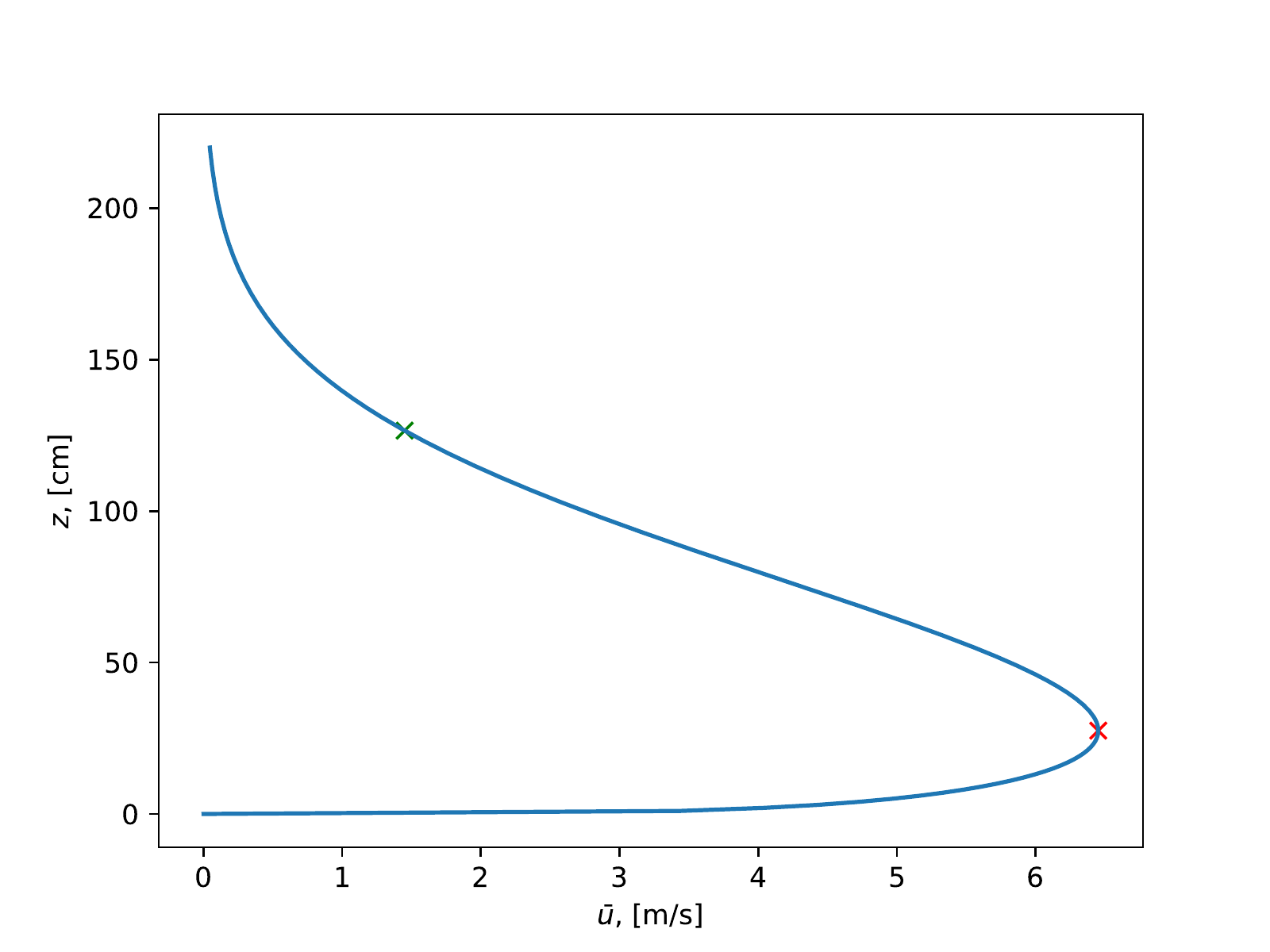}
	\caption{Plot of the time-averaged power-Gaussian velocity profile, as defined in Eq.~\eqref{eq:mean_velocity} with the parameters listed in Tab.~\ref{tab:velocityMean0000}. Red and green crosses are $z_0$ and $h$, respectively.}
	\label{fig:velocityMean0000}
\end{figure}

A simplification of the inlet field data comes from the assumption that inlet conditions are stationary. To obtain the optimal stationary profile, we fitted the data with constant $U,\, z_u\,, \xi\,, \chi$ (or, in other terms, we seek the 0th order polynomial fit for the four parameters). The results are reported in Tab.~\ref{tab:velocityMean0000} and Fig.~\ref{fig:velocityMean0000}.

\begin{table}[h!]
	\centering
\begin{tabular}{ll}
\toprule
{\bf Time-averaged, 2D/3D} & \\
\midrule
maximum velocity & $U(t) = U_0 = 6.451$ m/s\\
maximum velocity height & $z_u(t) = z_0 = 27.51$ cm\\
wall exponent & $\xi(t) = \xi_0 = 0.2374$\\
outer layer thickness & $\chi(t) = \chi_0 = 3.600$\\
\bottomrule
\end{tabular}
\caption{Fitting parameters for time-averaged (0,0,0,0 fit) power-Gaussian velocity profile defined by Eq.~\eqref{eq:mean_velocity}. }
\label{tab:velocityMean0000}
\end{table}

\subsubsection{Depth-averaged velocity data.}\label{sec:depth_average_evolution}
\begin{figure}[t!]
\centering
\includegraphics[width=\columnwidth]{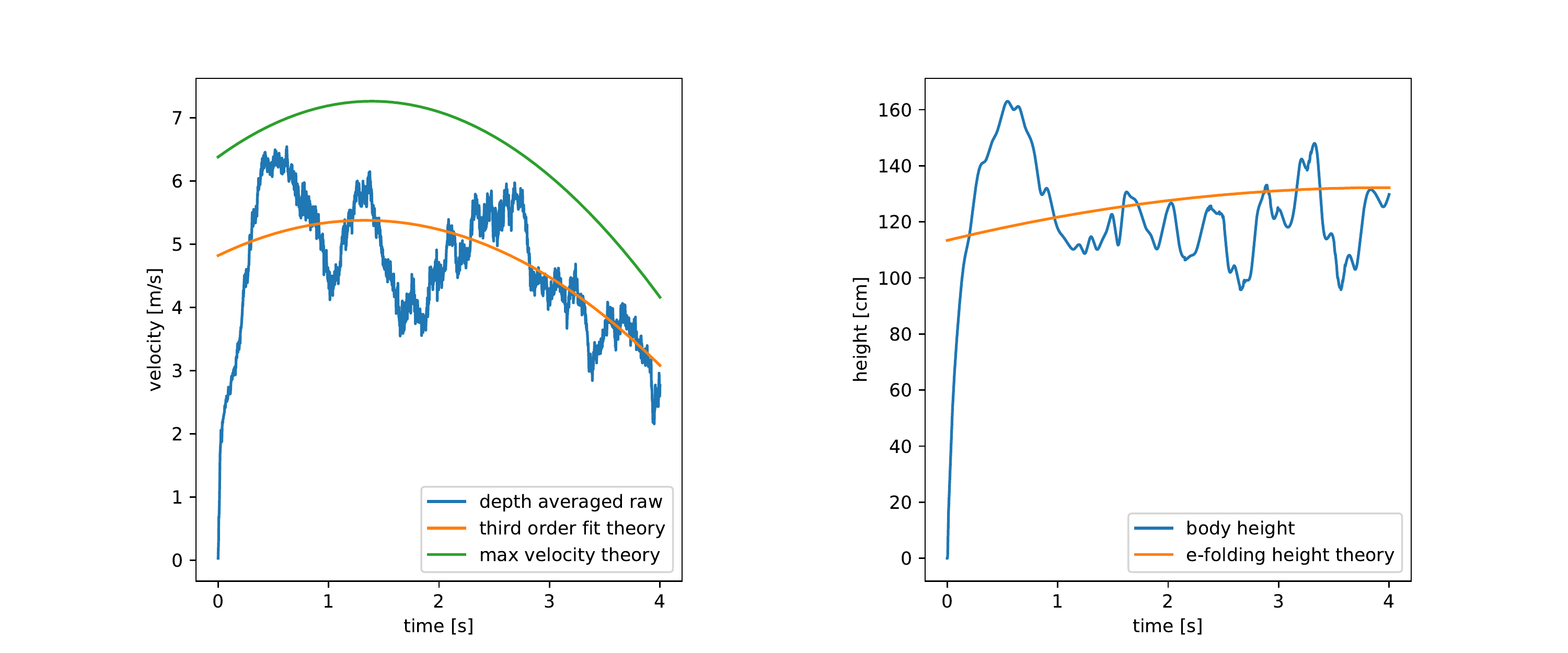}
\caption{Left panel: depth averaged velocity evolution based on the e-folding height $h(t)$; raw data are in blue, third order polynomial fit of the depth averaged theory is in orange. The green line is the maximum velocity $U(t)$. Right panel: the comparison between the body height $h_\mathrm{body}(t)$ and the height of the current $h(t)$.}
\label{fig:par_averages}
\end{figure}

To obtain the depth average velocity value, we use the e-folding height $h(t) = z_u(t) (\chi(t)+1)$ (see Tab.~\ref{tab:velocityMean}) as the flow thickness. This is comparable with the height of the current body as obtained manually from visual analysis of the particle concentration (see, e.g., Fig.~\ref{fig:par_averages}). 
In Fig.~\ref{fig:par_averages}, we compare the maximum velocity evolution with the depth average $\langle \bar{u} \rangle_z (t)$. This time series is compared with the depth averaged theory obtained integrating Eq.~\eqref{eq:mean_velocity}, which turns out to be a third order polynomial except for a 0.03\% of error.
Its coefficients are defined in Tab.~\ref{tab:par_averages}. 
$\langle \bar{u} \rangle_z (t)$ and $h(t)$ can be used in 1D depth averaged models. 
$\langle \bar{u} \rangle_z (t)$ can also be obtained as a fraction of the maximum velocity evolution $\langle \bar{u} \rangle_z (t) \simeq 0.741\,U(t)$ with a margin of error of 0.7 \%. 

\begin{table}[t!]
	\centering
	\begin{tabular}{cc}
		\toprule
		\multicolumn{2}{l}{\bf Transient, depth-averaged}\\
		\toprule
		velocity & $\langle \bar{u} \rangle_z (t) = u_0 + u_1 t + u_2 t^2 + u_3 t^3$\\
		$u_0$ & 4.818 m/s\\
		$u_1$ & 0.8409 m/s$^2$\\
		$u_2$ & -0.3160 m/s$^3$\\
		$u_3$ & -7.094e-4 m/s$^4$\\
		\midrule
		flow height & $h(t) = z_u(t) (\chi(t)+1)$\\ 
		\bottomrule
	\end{tabular}
	\caption{Parameters of the third order polynomial fit for depth-averaged inlet velocity. Temporal variation parameters for the flow height are reported in Tab.~\ref{tab:velocityMean} }
	\label{tab:par_averages}
\end{table}

\subsubsection{Time- and  depth-averaged velocity data.}
Finally, we have taken the time-average of the time series presented in
Sec.~\ref{sec:depth_average_evolution} to obtain a steady-state, depth-averaged
inlet condition. Using the parameters listed in Tab.~\ref{tab:velocityMean0000} to obtain the flow thickness $h_0$,
we have calculated the averaged velocity $\langle \bar{u} \rangle_{z,t}$ based $h_0$ . They are given in Tab.~\ref{tab:u_steady}.
\begin{table}
\centering
	\begin{tabular}{lc}
		\toprule
		\multicolumn{2}{l}{\bf Steady, depth-averaged} \\
		\toprule
		velocity & $\langle u \rangle_{z,t}$ = 4.80 m/s \\
		\midrule
		flow height & $h_0 = z_0(\chi_0 +1 )$ = 126.5 cm \\
		\bottomrule
	\end{tabular}
	\caption{Time- and depth-averaged values of inlet velocity and flow thickness}
	\label{tab:u_steady}
\end{table}

\subsection{Parallel velocity fluctuations.}
\begin{figure}[t!]
	\centering
	\includegraphics[width=\columnwidth]{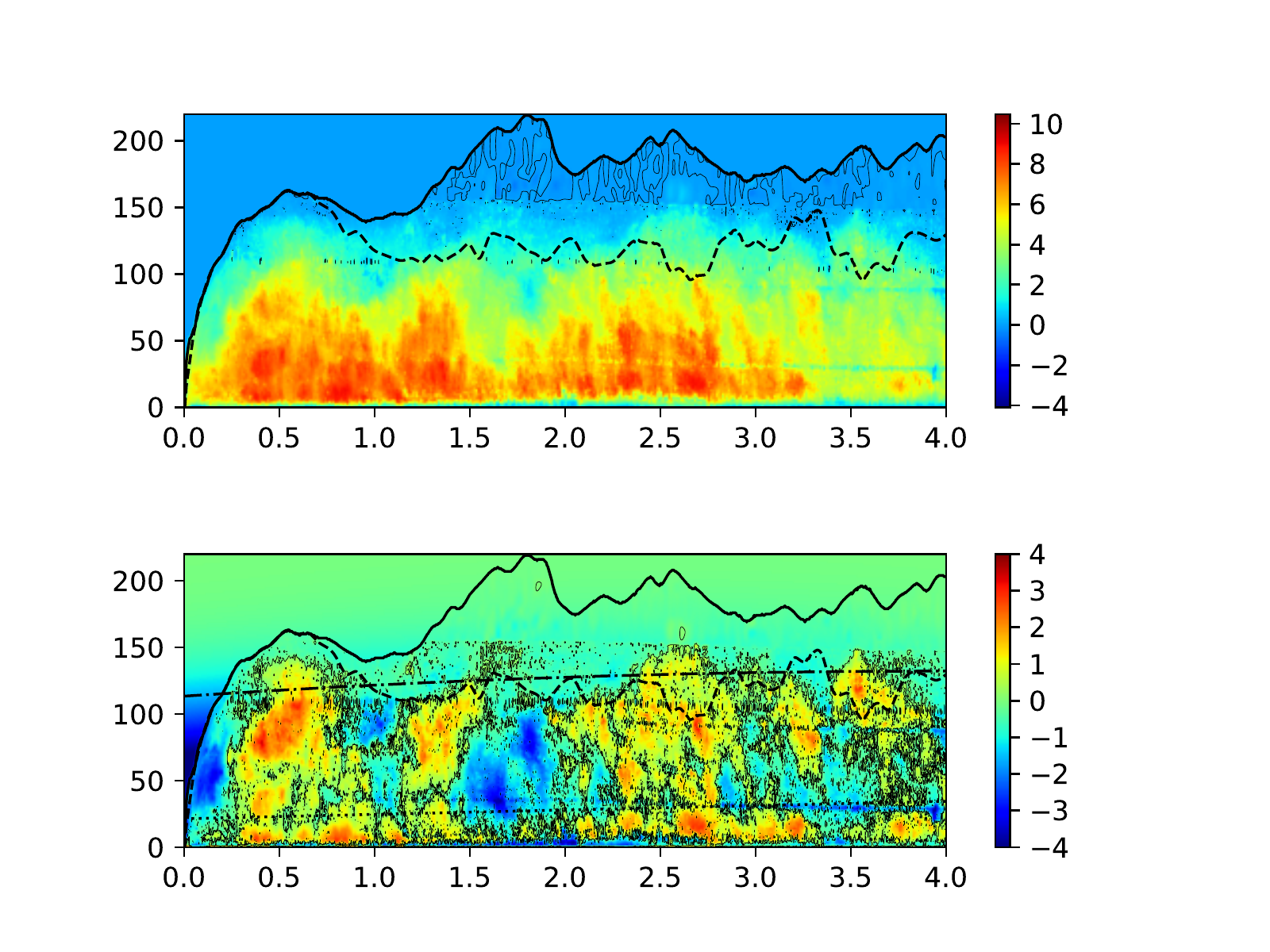}
	\caption{Parallel flow velocity. Upper panel, raw data (same as Fig. \ref{fig:par_mean}); lower panel, fluctuations $u'$. Color scale is in m/s; abscissa is time in seconds, ordinate is height in centimeters. Relevant heights are plotted as a function of time: $h_\mathrm{max}$ solid line; $h_\mathrm{body}$ dashed line; $h$ dash-dot line; $z_u$ dotted line.}
	\label{fig:par_fluct}
\end{figure}

Having defined a mean theory for the parallel velocity $u_\mathrm{pg}(z,t)$ (Eq. \ref{eq:mean_velocity}), it is possible to define velocity fluctuations from the raw data as $u'(z,t) = u(z,t) - u_\mathrm{pg}(z,t)$. 
In Fig.~\ref{fig:par_fluct} we display raw velocity data (upper panel) and the fluctuations (lower panel), obtained by subtracting the two plots of Fig.~\ref{fig:par_mean}.

\begin{figure}[t!]
\centering
\includegraphics[width=0.49\columnwidth]{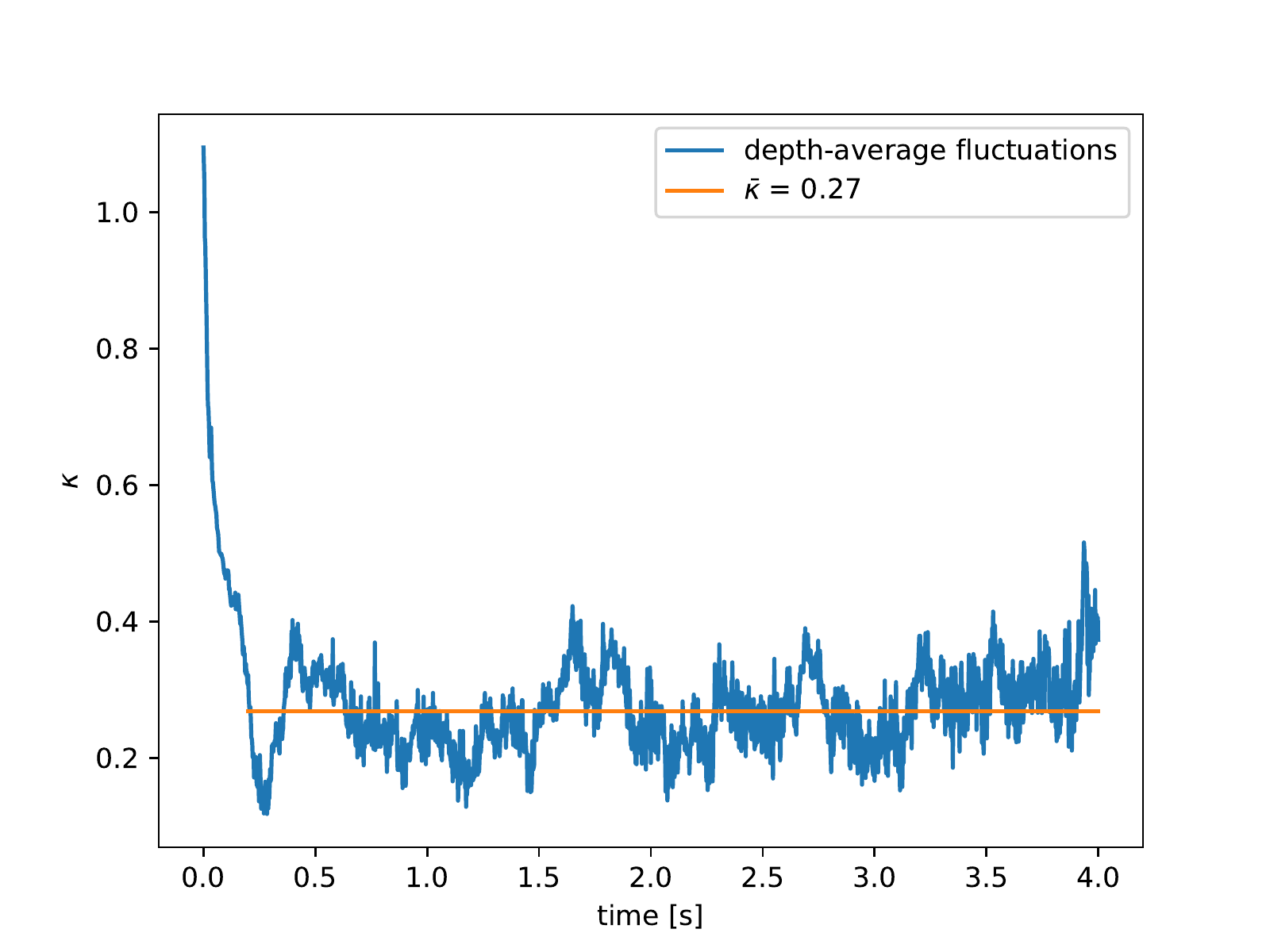}
\includegraphics[width=0.49\columnwidth]{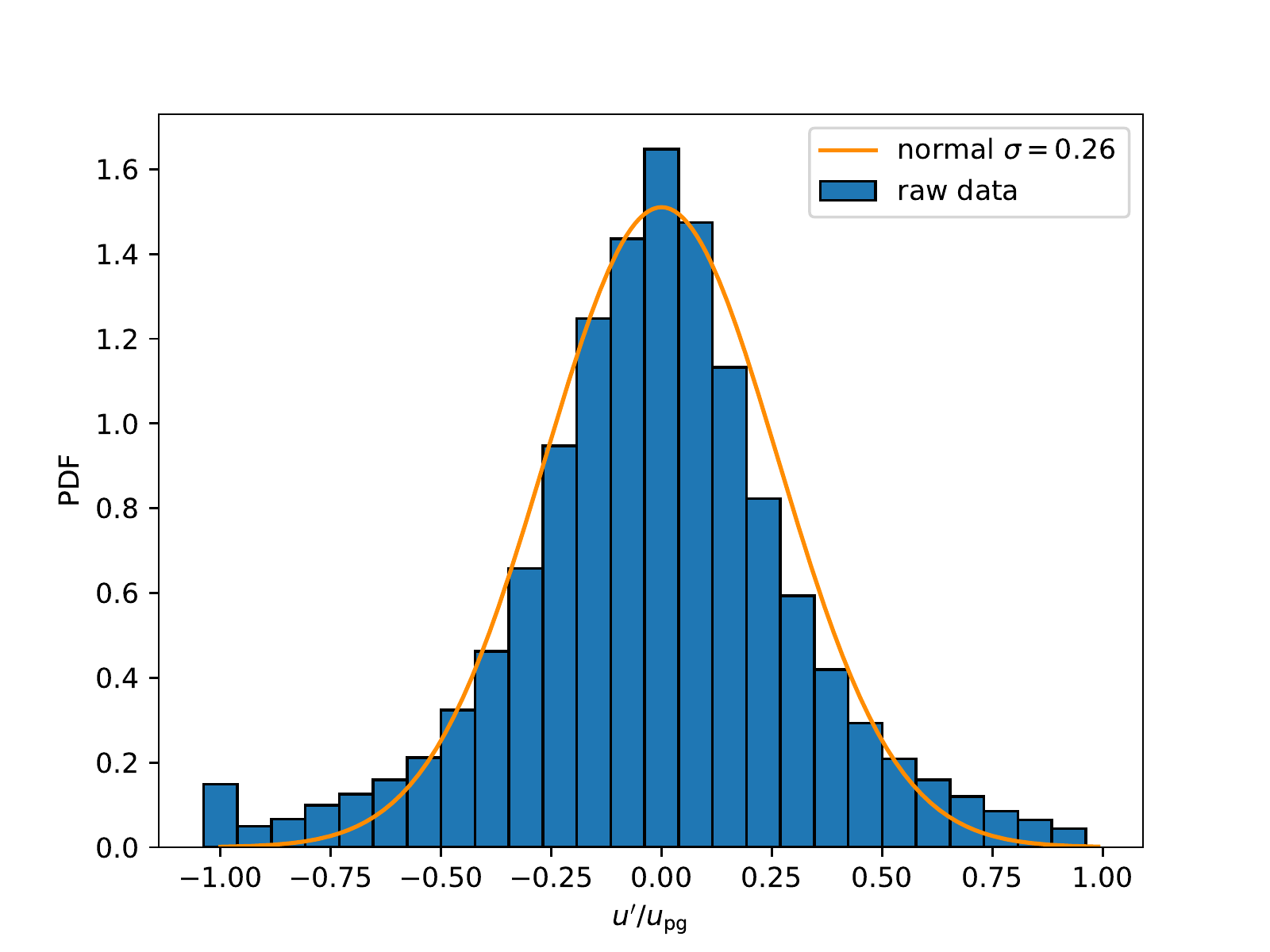}
\caption{Left panel: depth averaged fluctuation intensity with respect to the mean, up to $h_\mathrm{body}$; raw data are in blue, while the time average in the window $t = [0.2, 4]$ s is in orange. Right panel: probability density function of the fluctuation intensity up to $h_\mathrm{body}$ and in the window $t = [0.2, 4]$; raw data are in blue, while the fitting normal distribution with $\sigma \simeq 0.26$ is in orange.}
\label{fig:par_flt}
\end{figure}

We now provide a theory for the fluctuations. As for turbulent plumes and other shear flows (cf. \citet{Cerminara2015ashee}), we check whether turbulent fluctuations are proportional to the mean velocity profile. 
In Fig.~\ref{fig:par_flt}, we report the time series of the depth averaged fluctuations with respect to the depth averaged mean velocity:
\begin{equation}
\kappa (t) = \frac{\sqrt{\langle (u - u_\mathrm{pg})^2 \rangle}}{\langle u_\mathrm{pg}\rangle}\,.
\end{equation}

It is worth noting that, after the passage of the first part of the head, the magnitude of the relative fluctuations in the body of the current remains approximately constant, around a value $\bar{\kappa} = 0.27$. Moreover, we check that fluctuations distribute approximately in a normal way, with standard deviation $\sigma \simeq 0.26$, coherently similar to $\bar{\kappa}$. This result tells us that the the root mean square of velocity fluctuations is around 30\% the mean. This value is very similar to what has been found for plumes (cf. Fig.~14 in \citet{Cerminara2015ashee}).

\begin{figure}[t!]
\centering
\includegraphics[width=0.49\columnwidth]{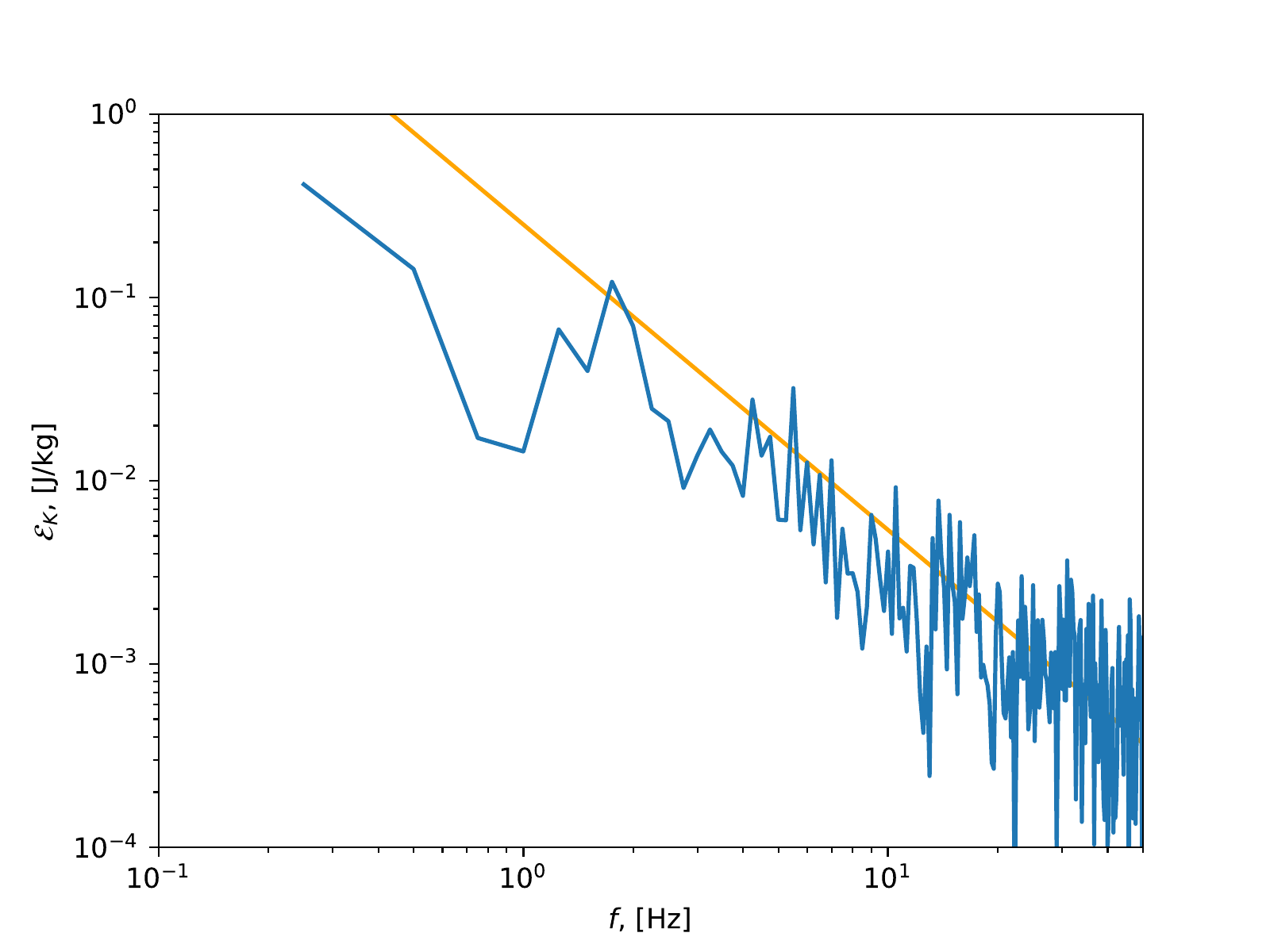}
\includegraphics[width=0.49\columnwidth]{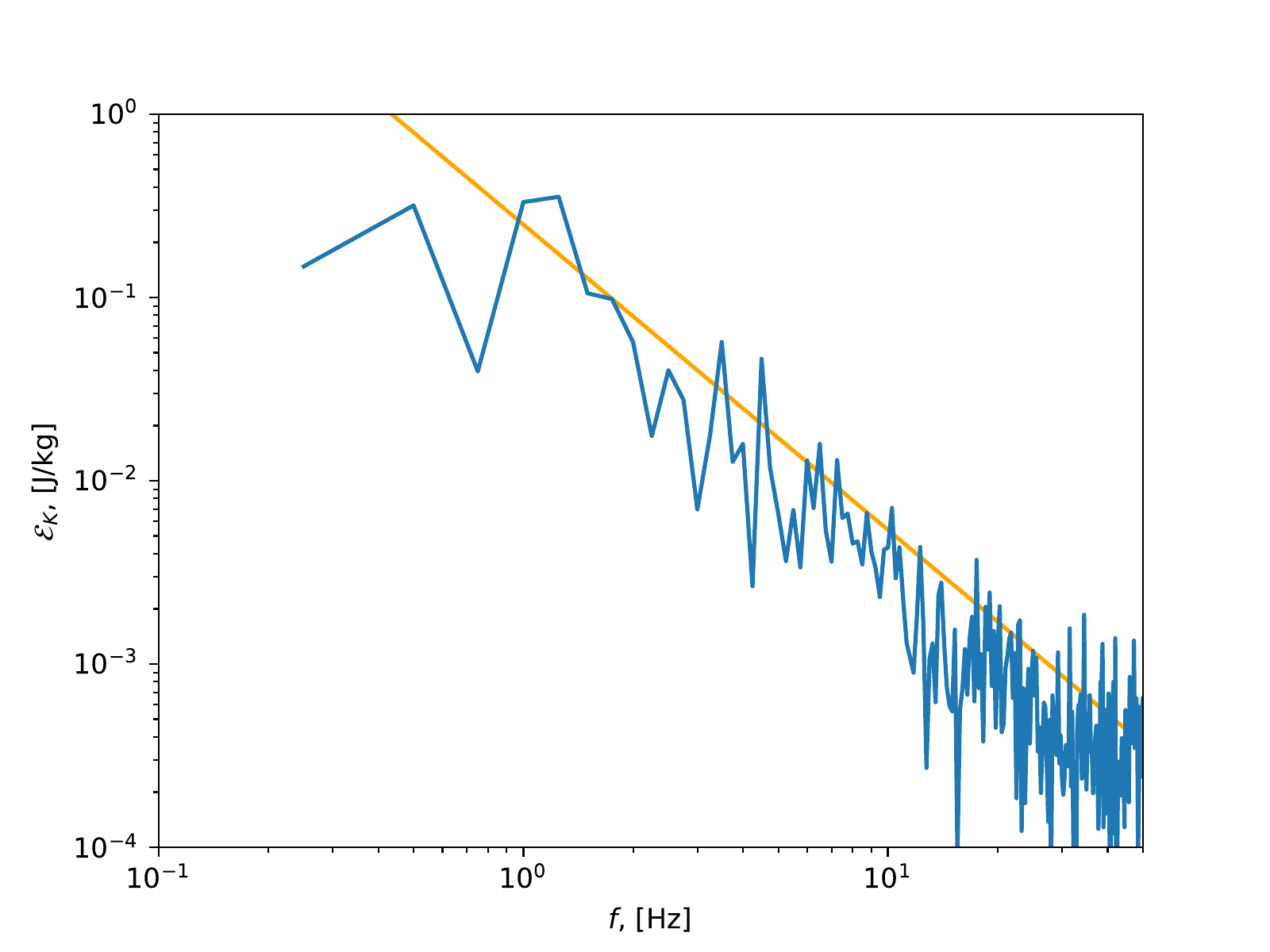}
\caption{Energy spectrum of the specific kinetic energy obtained from PIV data (blue lines). Orange lines are the Kolmogorov -5/3 turbulence cascade. Left panel: 15 cm above the bottom wall; right panel 80 cm.}
\label{fig:spectrum}
\end{figure}

PIV analysis allow us to confirm that turbulent fluctuations follow the classical Kolmogorov universal -5/3 cascade (see Fig.~\ref{fig:spectrum}). However, for simplicity, we choose to disregard any correlation in turbulent fluctuations, and model them as a simple white noise. We thus propose to approximate the velocity fluctuations as a fraction of the mean profile ($\simeq 30\%$), distributed normally around it. 

\subsection{Temperature and solid mass fractions.}\label{sect:temp}

\begin{figure}[t!]
\centering
\includegraphics[width=0.99\columnwidth]{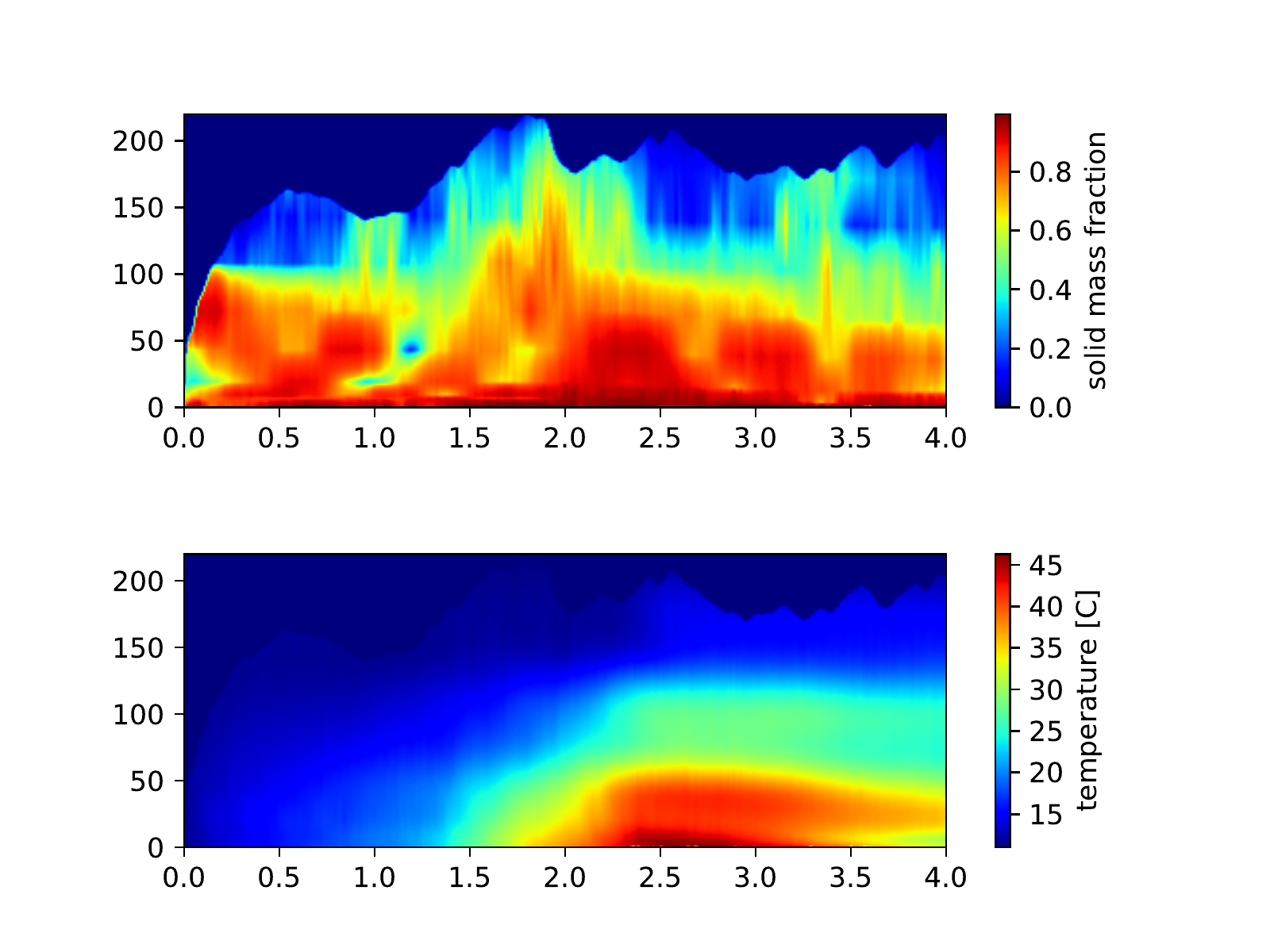}
\caption{Upper panel, solid mass fraction of solids in the flow. Lower panel, temperature field. Vertical axis is in cm, horizontal axis is in seconds.}
\label{fig:ys-T}
\end{figure}

The temperature and concentrations fields are obtained directly from the experiment using thermocouples and ash collectors placed at distances from the bottom wall reported in Table \ref{tab:probes}. An ambient temperature of $T_a = 11.0$ $^\circ$C is used to interpolate the temperature field outside the flow edge. Moreover, fixing the ambient pressure to $p_a = 1.013$ bar, humidity to 60 \%, the gas constant of air and water respectively to 287 and 462~J/(kg K), the mass fraction of the solid phase can be calculated. Resulting data are shown in Fig.~\ref{fig:ys-T}.

\subsubsection{The theoretical model for temperature and concentration.}
We simplify both the temperature and mass fraction profiles by neglecting their temporal variability, and give their time-average profile. Data are fitted with the theoretical exponential profile proposed by \citet{Cantero-Chinchilla2015} for gravity currents:
\begin{equation}\label{eq:cantero}
f(\eta) = F \exp\left(-\phi \eta^\zeta\right)\,,
\end{equation}
where $F$ is the value in $\eta = z/z_u = 0$, $\phi$ and $\zeta$ are two non-dimensional parameters. This profile law can be used both for the temperature $T$ and solid mass fraction $y_\mathrm{s}$ distribution. 

\subsubsection{Time averaged data of temperature and concentrations.}
\label{sec:time_averaged_profiles_ys-T}
\begin{figure}[t!]
\centering
\includegraphics[width=0.49\columnwidth]{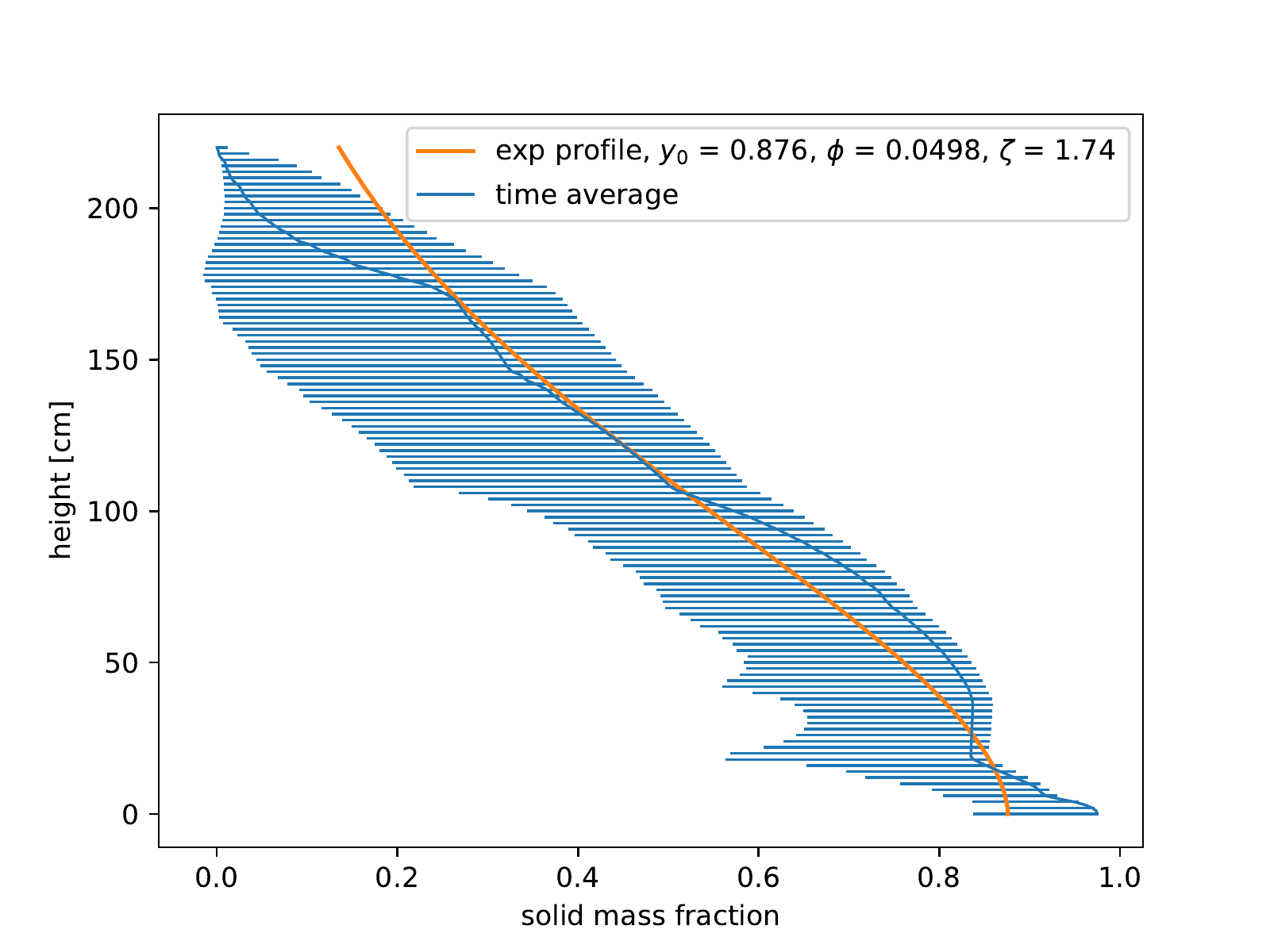}
\includegraphics[width=0.49\columnwidth]{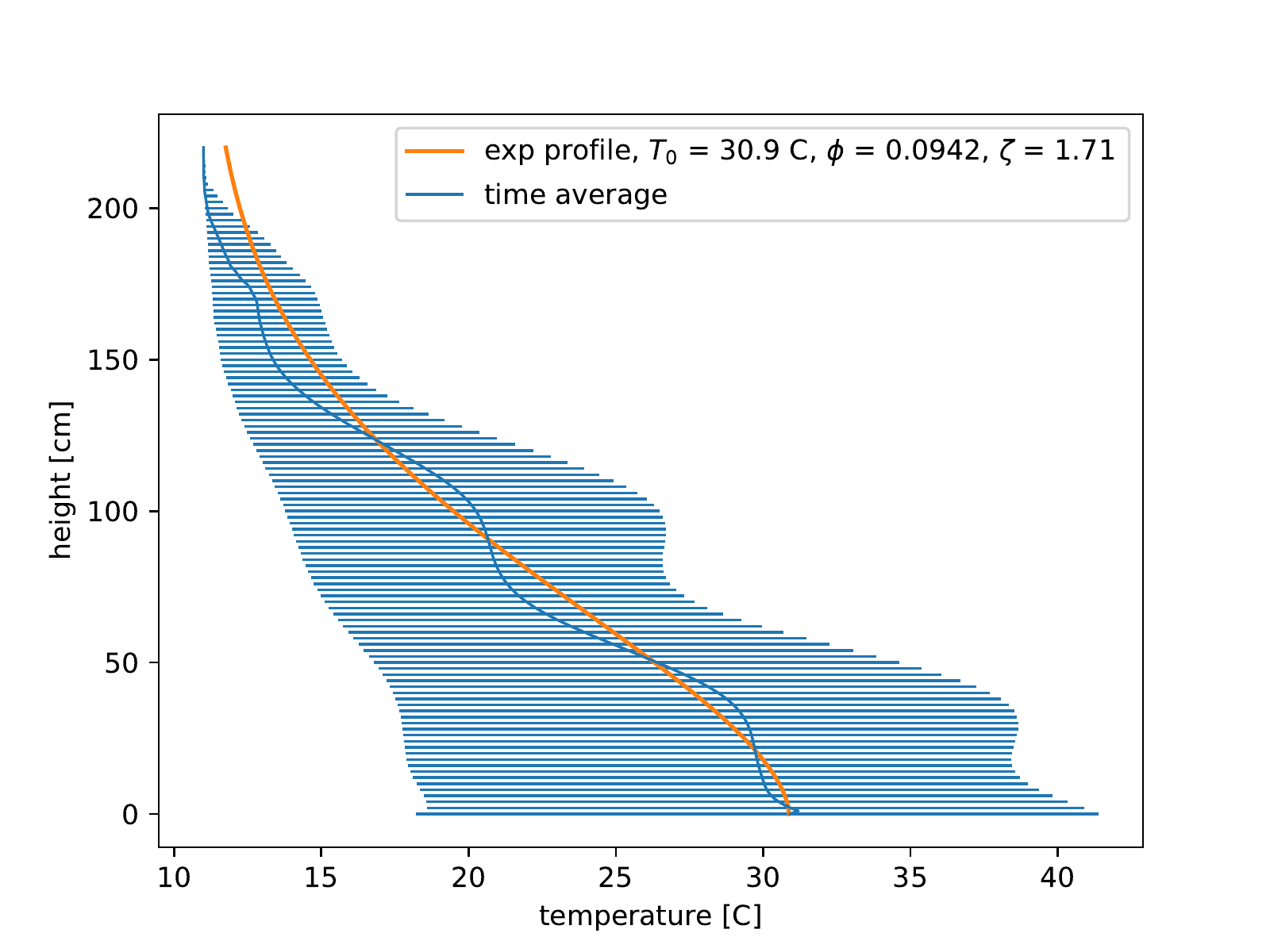}
\caption{Left panel, solid mass fraction; right panel, temperature. Blue line is the time average with error bars, the orange line is the fit obtained with eq.~\eqref{eq:cantero} and the parameters reported in legend and in Tab.~\ref{tab:ys-T}.}
\label{fig:ys-T-mean}
\end{figure}

Results of the fitting procedure are reported in Tab.~\ref{tab:ys-T} and Fig.~\ref{fig:ys-T-mean}, using \mbox{$z_u = 27.51$ cm}, as for the time-averaged velocity profile.

\begin{table}[h!]
\centering
\begin{tabular}{ccc}
\toprule
\multicolumn{3}{l}{{\bf Time-averaged, 2D/3D}}\\
\midrule
 & solid mass fraction & temperature\\
 \midrule
bottom value $F$ & $y_0 = 0.876$ & $T_0 = 30.9$ $^\circ$C \\
$\phi$ & 0.0498 & 0.0942 \\
$\zeta$ & 1.74 & 1.71 \\
\bottomrule
\end{tabular}
\caption{Parameters for the time-averaged profiles of solid mass fraction and temperature, obtained with eq.~\eqref{eq:cantero} and $z_u = 27.51$~cm.}
\label{tab:ys-T}
\end{table}

\subsubsection{Depth-averaged temperature and concentration data.}
\label{sec:depth_average_evolution_ys-T}
\begin{figure}
\centering
\includegraphics[width=0.49\columnwidth]{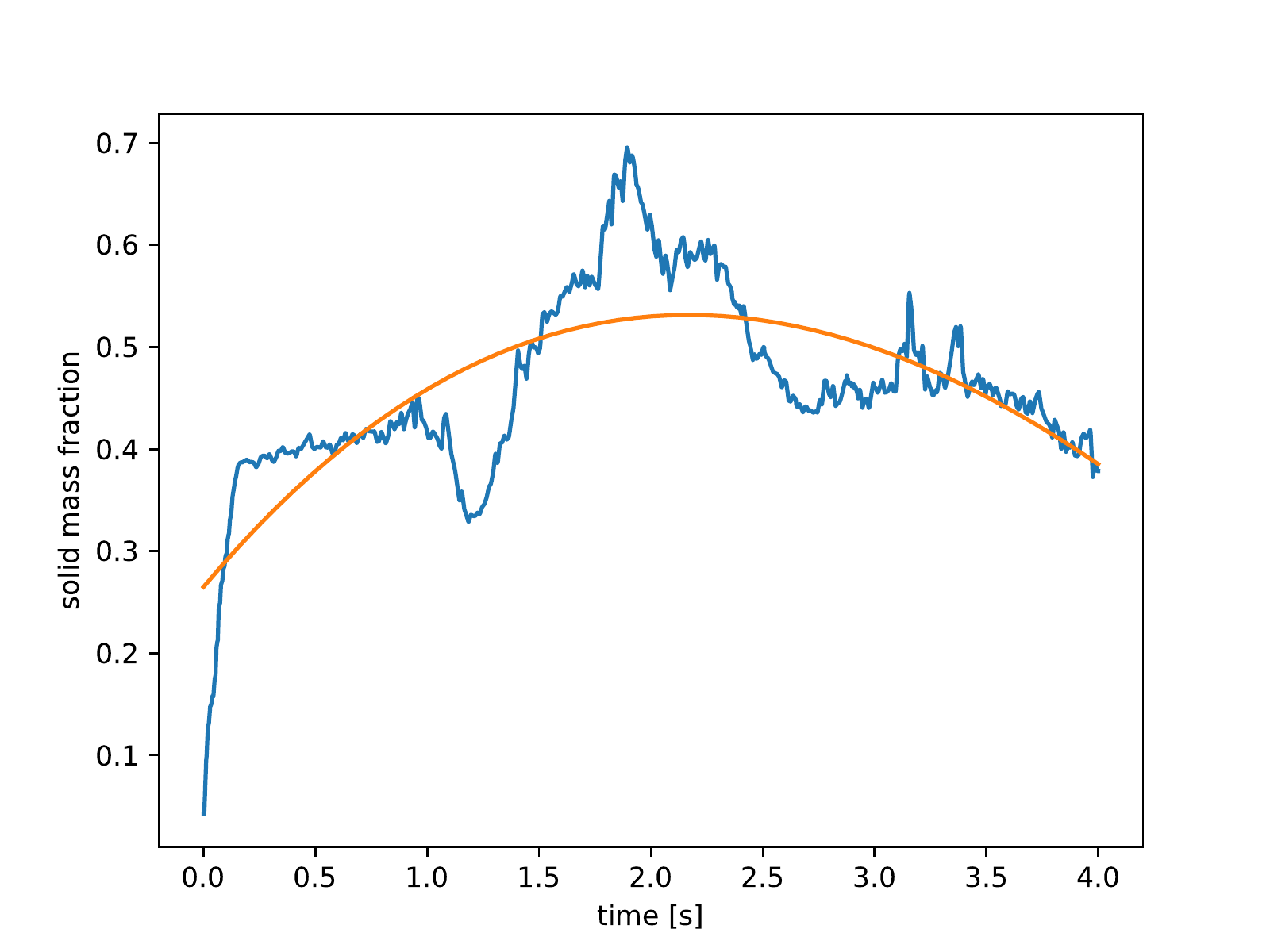}
\includegraphics[width=0.49\columnwidth]{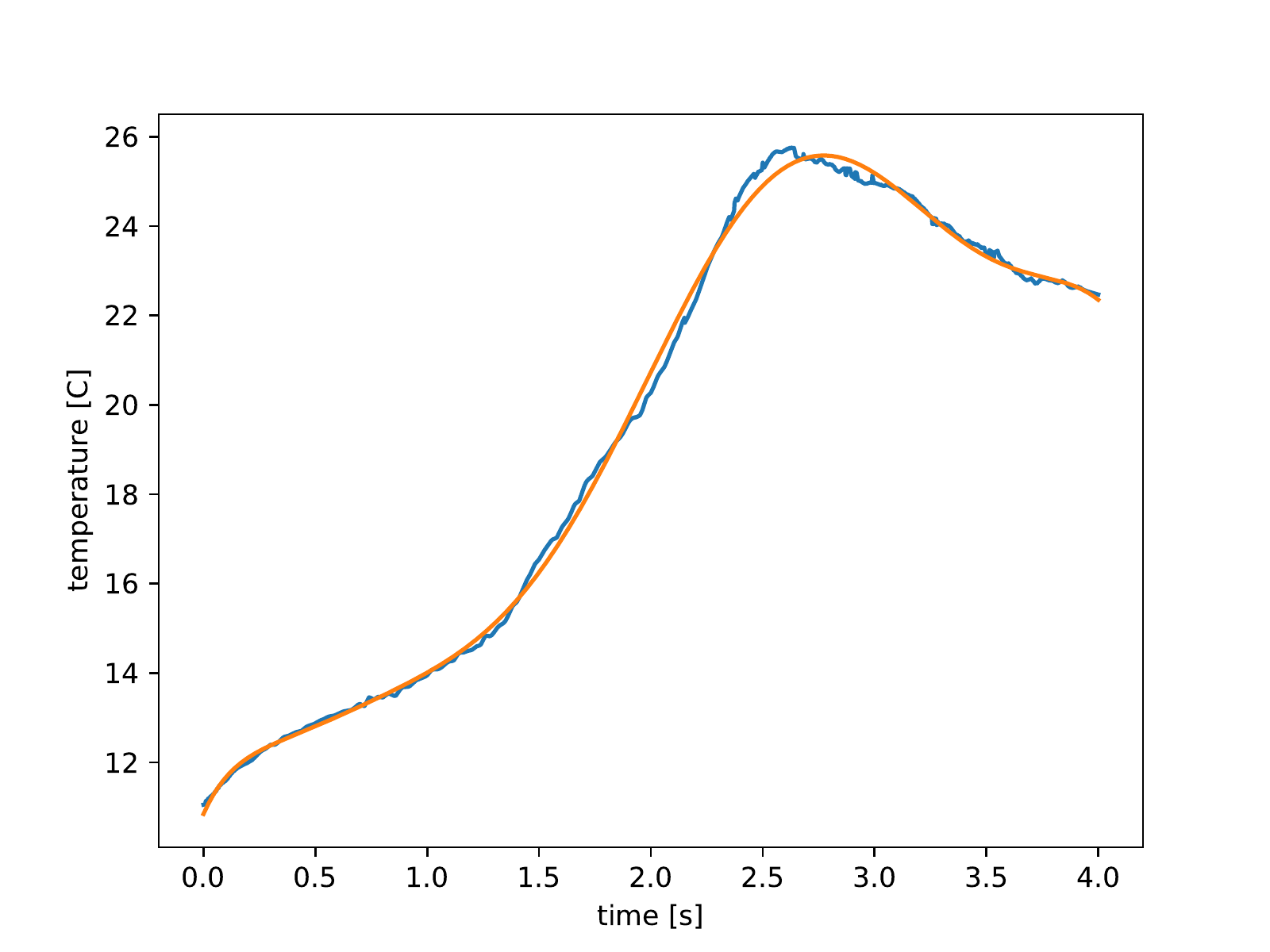}
\caption{Depth averaged data (blue lines) and polynomial fit (orange line) for solid mass fraction (left panel) and temperature (right panel).}
\label{fig:ys-T_averages}
\end{figure}

As done in Sec.~\ref{sec:depth_average_evolution}, depth-averaging of the temperature and solid mass fraction fields can be done to produce transient boundary conditions for depth-averaged models. 
The model chosen for the time evolution of these two fields is again polynomial:
\begin{equation}
p(t) = \sum_{n=0}^N p_n t^n
\end{equation}
with $N=3$ for the solid mass fraction and $N=9$ for the temperature.
While for the mass fraction of solids a $3^{\rm rd}$ order polynomial is enough to obtain a satisfactorily fit (net of fluctuations, as done for the velocity in Fig.~\ref{fig:par_averages}), the temperature evolution is more complicated, without oscillating fluctuations. Thus, we decided to follow strictly its evolution, with a $9^{\rm th}$-order polynomial fit. Results are shown in Fig.~\ref{fig:ys-T_averages}, while fitting parameters are reported in Tab.~\ref{tab:ys-T_averages}.
%
%
\begin{table}
\centering
\begin{tabular}{lcc}
\toprule
\multicolumn{3}{l}{\bf \normalsize Transient, depth-averaged}\\
\midrule
& $y_\mathrm{s}$ & $T$\\
\midrule
$p_0$ & 0.26496133 & 10.8459539\\
$p_1$ & 0.26126439 & 11.1623744\\
$p_2$ & -0.0709373 & -36.2955379\\
$p_3$ & 0.00328635 & 76.0021061\\
$p_4$ & - & -91.0330589\\
$p_5$ & - & 64.3634994\\
$p_6$ & - & -26.5272360\\
$p_7$ & - & 6.21846488\\
$p_8$ & - & -0.767026878\\
$p_9$ & - & 0.0386039457\\
\bottomrule
\end{tabular}
\caption{Coefficients used for the polynomial fits in Fig.~\ref{fig:ys-T_averages}, with dimensions such that temperature is in Celsius and time in seconds.}
\label{tab:ys-T_averages}
\end{table}

\subsubsection{Time- and depth-averaged temperature and concentration data.}
To conclude this section on solid mass fraction and temperature fields, their
scalar value, averaged both in time and space should be given. Again, we take
the time-average of the time series presented in the previous section, to
obtain values reported in Tab.~\ref{tab:ys-T_steady}
\begin{table}
\centering
\begin{tabular}{lc}
	\toprule
	{\bf Steady, depth-averaged} & \\
	\midrule
	$\langle y_\mathrm{s} \rangle$ & 0.4617  \\
	$\langle T \rangle$ & 19.29 $^\circ$C\\
	\bottomrule
\end{tabular}
\caption{Steady, depth-averaged values of inlet particle mass fraction and temperature}
\label{tab:ys-T_steady}
\end{table}

\subsection{Grain-size distribution}\label{sect:gsd}
\begin{figure}[h!]
\centering
\includegraphics[width=0.49\columnwidth]{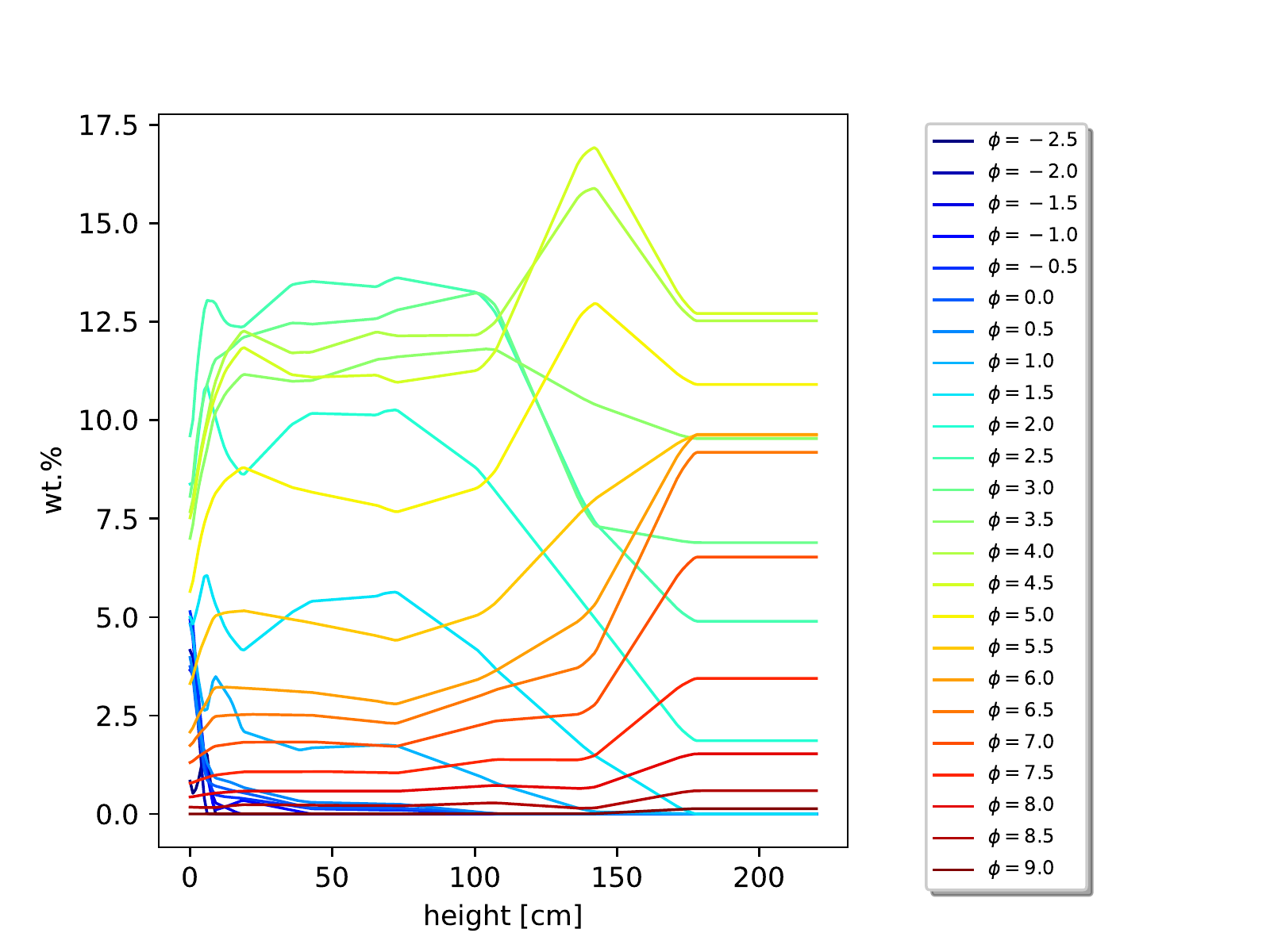}
\includegraphics[width=0.49\columnwidth]{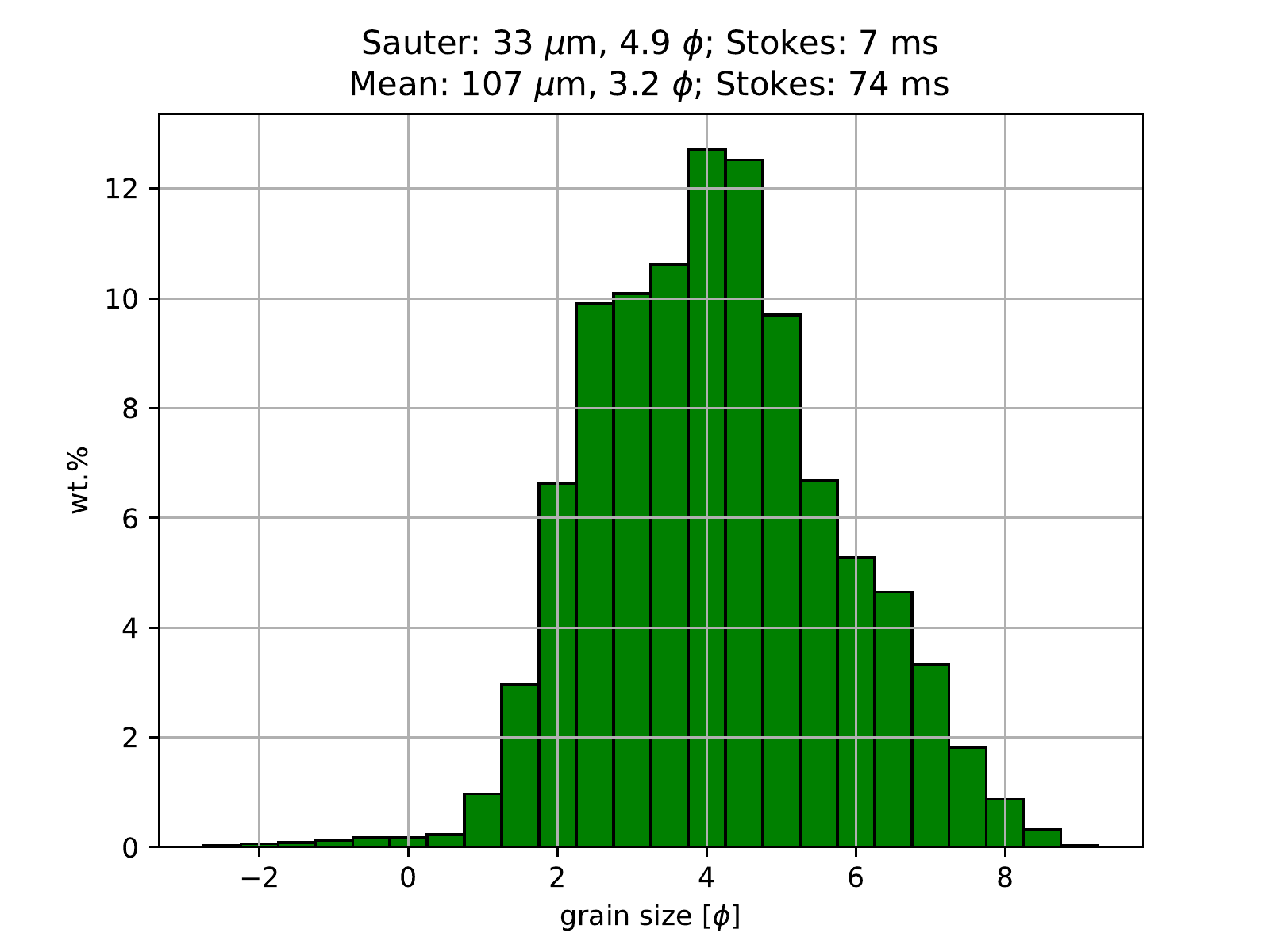}
\caption{Grain-size distribution (weight percent), averaged in time (left panel) and averaged both in time and space (right panel). The size of the particle classes is given in $\phi$ units. The Sauter and mean diameters of the time and space averaged distribution are on top of the right panel, with their Stokes times calculated with the particle density (see Tab.~\ref{tab:gsd}) and the viscosity of air in standard conditions.}
\label{fig:gsd}
\end{figure}

The time and space averages of the grain-size distribution (GSD) are reported in Fig.~\ref{fig:gsd}. The GSD is measured in the same locations where the concentration is measured, by sieving the ash deposition history at different heights. The mass fraction can thus be interpolated to obtain a time-space field for each $j$-th particle class $y_j(z, t)$. 
We report in Fig.~\ref{fig:gsd} (left panel) the time average of mass fraction $y_j$ for each particle class. Although the flow appears to be already stratified and particle concentration varies with depth, adding such a complicated data set into the benchmark is beyond the scope of the present work. We thus use the time-space average of the GSD, to obtain a standard distribution that does not depend on time and space (shown in Fig.~\ref{fig:gsd}, right panel), strongly simplifying the boundary conditions for the benchmark.
\begin{table}
\begin{scriptsize}
\begin{tabular}{lcccccccccccc}
\toprule
diameter [$\phi$] & -2.5 & -2. & -1.5 & -1 & -0.5 & 0 & 0.5 & 1 & 1.5 & 2 & 2.5 & 3\\  
wt.\% &0.04 & 0.07 & 0.09 & 0.12 & 0.17 & 0.18 & 0.23 & 0.97 & 2.96 & 6.63 & 9.91 & 10.09\\
density [kg/m$^3$] & 510 & 620 & 739.9 & 859.7 & 1074 & 1289 & 1480 & 1672 & 1824 & 1975 & 2085 & 2194\\
\midrule
diameter [$\phi$] & 3.5 & 4 & 4.5 & 5 & 5.5 & 6 & 6.5 & 7 & 7.5 & 8 & 8.5 & 9\\
wt.\% & 10.61 & 12.72 & 12.52 & 9.70 & 6.68 & 5.28 & 4.65 & 3.32 & 1.82 & 0.88 & 0.32 & 0.04\\
density [kg/m$^3$] & 2369 & 2544 & 2572 & 2600 & 2600 & 2600 & 2600 & 2600 & 2600 & 2600 & 2600 & 2600\\
\bottomrule
\end{tabular}
\end{scriptsize}
\caption{Grain-size distribution at the inlet: time and space averaged data. }
\label{tab:gsd}
\end{table}

\section{Post-processing of numerical results}\label{sect:postproc}


 











\subsection{Strategy}
The strategy for comparing experimental and numerical models outputs is similar to that put in place for setting up the initial conditions. Comparison will be carried out at four vertical sections (Profiles 2,3,4 and 5 in Figure 2), where experimental and numerical profiles will be analyzed in their instantaneous, time-averaged and depth-averaged values. 
The most complete comparison would be obtained with numerical data sampled at 100~Hz (in particular, high-frequency data will be used to calculate fluctuations statistics and energy spectra). If is a problem to save the data at 100~Hz, a high-frequency sampling at given probe locations is required. In particular, at each profile section at the heights above the bottom surface listed in Tab.~\ref{tab:probes}.
\begin{table}[t!]
\centering
	\begin{tabular}{lccccccccc}
		\toprule
		Probe & $z_1$ & $z_2$ & $z_3$ & $z_4$ & $z_5$ & $z_6$ & $z_7$ & $z_8$ & $z_9$\\
		\midrule
		Distance [m] & 0.035 & 0.08 & 0.11 & 0.21 & 0.45 & 0.75 & 1.1 & 1.4 & 1.8\\
	\bottomrule
	\end{tabular}
\caption{Distances (orthogonal to bottom plane) of temperature and concentration probes at sampling positions.}
\label{tab:probes}
\end{table}

Along the profiles or at the probe locations, the following field data will be compared (if available from the model):
\begin{itemize}
	\item Velocity
	\item Temperature
	\item Mass Fraction (of each particle class)
	\item Grain Size Distribution
\end{itemize}
In addition to the above fields, front position,  average flow thickness (and their time wise evolution), and final deposit thickness will be compared.
Because models at different levels of approximation will not provide the same of information, experimental data will be filtered accordingly for carrying out the comparison (as described in Section~\ref{sect:bc}).

\subsection{Inter-Comparison methodology}
Plots will be elaborated using a standardized procedure. Participants are asked to provide data at Profiles 2-4 in a common format (Sect. \ref{sect:dataformat})

Data will be processed to produce the following outputs:
\begin{itemize}
	\item For all models, the time-dependent run-out and thickness plot. 
	\item If available, the final deposit and the deposit evolution at the sampling distances.
	\item Contour plots for velocity, temperature and mass fraction of solids (as in Figs. \ref{fig:par_mean} and \ref{fig:ys-T}) for 2D and 3D time-dependent models.  
	\item The fit of velocity data with the theoretical velocity profile of Eq.~\ref{eq:mean_velocity} (as in Fig.~\ref{fig:par_pgt}) for 2D and 3D time-dependent models.
	\item Time-averaged flow velocity, temperature, mass fraction of solids, and grain size distribution profiles (as in Fig.~\ref{fig:velocityMean0000},~\ref{fig:ys-T-mean}, and~\ref{fig:gsd}a) for 2D and 3D models.
	\item Time-dependent, depth-averaged values (as in Fig.~\ref{fig:par_averages} and~\ref{fig:ys-T_averages}) and Time- and Depth- averaged plots (as in Tab.~\ref{tab:u_steady},~\ref{tab:ys-T_steady}, and~\ref{tab:gsd}) for all models.
	\item Velocity fluctuation statistics (as in Fig.~\ref{fig:par_fluct},~\ref{fig:par_flt}, and~\ref{fig:spectrum}) only for 2D and 3D time-dependent models (and if high-frequency data will be available) .
\end{itemize}

\subsection{Data format}\label{sect:dataformat}
Output data provided by numerical modelers should be organized in a standardized format, in order to be post-processed with the same application.
For every field (velocity, temperature, concentration of each grain, pressure), an individual file  should be provided for each profile distance (organized in folders).
For 3D and 2D models, every file should contain data along the vertical profile as an ASCII matrix.
Every row in the matrix should correspond to a different time, whereas columns correspond to different vertical position. The first row should contain z coordinates (in meters), the first column should report the time (in seconds after the beginning of the simulation, i.e., after flow arrival at Profile 1, the input location). Locations and times in the matrix where the flow is not present should be filled by zeroes.
For depth-averaged models, a single file should be provided for each profile position, where the first column is time and the other columns report the time-dependent values.

\section{Conclusion}\label{sect:concl}
We have provided inlet boundary conditions for comparing models at different levels of approximations with the experimental data obtained from the PELE large-scale experiment. This will provide the basis for defining an experimental benchmark for PDC models.
\begin{itemize}
\item 2D/3D transient models can impose transient, stratified mean profile of velocity, temperature and concentration at the inlet, as given by Eq.~\ref{eq:mean_velocity}~and~\ref{eq:cantero}, with coefficients given in Tab.~\ref{tab:velocityMean}~and~\ref{tab:ys-T}. While mean velocity profile can imposed to be either time-dependent or steady, concentration and temperature profiles are constant in time.
\item 1D depth-averaged models (or 2D/3D models willing to simplify the inlet conditions) can use the transient depth-averaged values provided in Tab.~\ref{tab:par_averages}~and~\ref{tab:ys-T_averages}.
\item Steady-state integral models can use the time- and depth-averaged values provided in Tab.~\ref{tab:u_steady}~and~\ref{tab:ys-T_steady}.
\item Velocity fluctuations can be imposed at the inlet by adding to the time-variant or mean parallel component of the velocity profile a Gaussian white noise of 30\% intensity with respect to the instantaneous velocity. Fluctuations can be also added to the depth-averaged and time-averaged inlet velocity.
\item Temperature and concentration fluctuations will be neglected, as they are not easily constrained by the experimental data.
\item Grain size distribution at the inlet is assumed to be constant in time and across the thickness. Averaged numerical values in $\phi$ scale are reported in Tab.~\ref{fig:gsd}.
\end{itemize}
Ambient initial conditions are reported in Tab.~\ref{tab:ambient}.

\section*{Acknowledgments}
This study was supported by the Royal Society of New Zealand Marsden Fund (contract no. MAU1506 and MAU1902), the New Zealand Natural Hazards Research Platform (contract no. 2015-MAU-02-NHRP) and the Resilience to Nature's Challenges National Science Challenge Fund New Zealand (GNS-RNC047). We thank Tomaso Esposti Ongaro for reviewing the manuscript.

\addcontentsline{toc}{chapter}{\refname}
\bibliographystyle{apalike}
\bibliography{article}


\end{document}